\documentclass[a4paper]{jpconf}
\usepackage{graphicx}

\def\br{\begin{eqnarray}}
\def\er{\end{eqnarray}}
\def\be{\begin{equation}}
\def\ee{\end{equation}}
\def\nonu{\nonumber}
\def\lb{\lbrack}
\def\rb{\rbrack}
\def\({\left(}
\def\){\right)}

\relax

\def\a{\alpha}
\def\b{\beta}

\def\d{\delta}

\def\k{\kappa}
\def\l{\lambda}

\def\o{\over}

\def\pa{\partial}

\def\lie{{\cal G}}
\begin{document}


\title{Backlund Transformation for   Integrable Hierarchies: example -  mKdV Hierarchy }

\author{J.F. Gomes, A.L. Retore, N.I. Spano and A.H. Zimerman}

\address{Instituto de F\'\i sica Te\'orica IFT-Unesp\\
S\~ao Paulo - Brasil}

\ead{jfg@ift.unesp.br, retore@ift.unesp.br, natyspano@ift.unesp.br, zimerman@ift.unesp.br}

\begin{abstract}
In this note  we present explicitly the construction of the mKdV hierarchy and show that it decomposes into positive and negative graded sub-hierarchies.  We extend the construction of the Backlund transformation for the sinh-Gordon model to all other  positive and negative odd graded equations of motion generated by the same affine algebraic structure. Some simple examples of solutions    are explicitly verified to satisfy, in a universal manner,  the Backlund transformations for the first few odd (positive and negative) sub-hierarchies.

\end{abstract}

\section{Introduction}

A subclass of non linear integrable models underlined by an affine ${\hat {sl}(2)}$ Lie algebra is well known to be connected to  the mKdV equation.  These are in fact higher flows obtained from the zero curvature representation \cite{nissimov} and by general algebraic arguments  these flows are restricted to be related to certain  positive odd grade generators.  Another subclass of nonlinear integrable models containing for instance, the sinh-Gordon model may be formulated, also within the zero curvature representation, but now associated to negative  grade generators.   The relation between the mKdV and sine-Gordon models  was already observed some time ago \cite{1}, \cite{2} in terms of conservation laws and by more algebraic arguments  it was generalized for the AKNS hierarchy in \cite{jpa} and to other integrable hierarchies  associated to  mixed gradations  allowing internal degrees of freedom in \cite{sotkov}, \cite{iraida} and  \cite{2bh}.  The general construction and classification  of the hierarchy  maybe understood  in terms of  a decomposition of the affine Lie algebra into graded subspaces  by a judicious choice of a grading operator.  The hierarchy is further specified  by choosing a constant grade one operator.  These  algebraic ingredients  define  a series of non-linear equations of motion, each corresponding to a different time evolution and hence to a hamiltonian structure (see for instance  \cite{nos} and references therein).  

Backlund transformation has recently been employed to extend the set of integrable  models to incorporate defects.  Such defects preserve the integrability when they are described by Backlund transformations connecting two distinct solutions of the same equation of motion at its location.  This was firstly observed in \cite{corrigan1} for the sine-Gordon  and extended to affine Toda field theories with defects  in \cite{corrigan-toda} and  to other non-relativistic models \cite{corrigan-nr}.  In this last reference  the same space component of the Backlund transformation  of the    sine-Gordon models is employed to describe integrable  defects within the mKdV model.  It thus seem natural the extend the same space component of the Backlund transformation to other members of the integrable hierarchy.

In this note we start by reviewing, in section 2, the algebraic construction for general  integrable hierarchies.  We discuss explicitly the mKdV Hierarchy constructed out of the affine $\hat {sl}(2)$ Kac-Moody algebra and  principal gradation.   We show how the {\it positive and negative odd} sub-hierarchis naturally arises from the zero curvature representation.  An important point to notice is that, while the time component  for the construction of each model varies with the gradation, the form of the space component remains the same for all models within the hierarchy.
Moreover, we  discuss, in section 3,  the construction of Backlund transformation  in terms of gauge transformation that preserves the form of the space Lax operator. By general arguments, the space Backlund transformation  is   understood to be universal     within the  hierarchy.  This fact  can be seen in  ref. \cite{wadati1} where  the space Backlund transformation for the sinh-Gordon and mKdV  were derived by canonical transformations and shown to agree.
Also, by the same procedure  the space Backlund transformation for the KdV    and the Sawada-Kotera equations ( belonging to the same hierarchy) obtained in \cite{wadati1} and \cite{villani} respectively agree.  
 Its time components can be extended to the whole hierarchy      by using the appropriated equations of motion. As an explicit example, we derive the space Backlund transformation for the sinh-Gordon model
 and construct its time components o for higher positive and negative grade time evolutions.

This, in fact  maybe observed by the explicit space-time dependence of the soliton solutions.   We  explicit display  the 1- and 2-soliton solutions for all (odd   positive and negative grade) models within the hierarchy and show  how they can be arranged in pairs in order to satisfy the Backlund  relations.  Some explicit examples are verified in section 4.

\section{The mKdV Hierarchy}

The main ingredient  underlying  the construction of  integrable hierarchies (IH) is the   Lax operator,
\br
L_x = \pa_x + E^{(1)} + A_0
\label{l1}
\er 
where $E^{(1)} $ and $A_0 $ are Lie algebra $\lie $ valued  elements and  carry an affine    structure which classifies the IH.  

The systematic construction of IH and its Lax operators  of the form (\ref{l1}) consists  in the decomposition of an affine algebra $\hat {\lie}$ into integer graded subspaces
\br
\hat \lie = \sum_{a} \lie_a, \;\;\;\;a\in Z
\er
 induced  by a choice of  a grading operator $Q$, such that
\br 
\lb Q, \lie_a \rb =a\lie_a, \qquad  \lb \lie_a , \lie_b \rb \in \lie_{a+b}, \qquad  a,b \in Z.
\er
Furthermore, the IH is determined by fixing the semi simple grade one operator $E^{(1)} \in \lie_1$ such that it decomposes $\hat \lie ={\cal K} \oplus {\cal M}$ where $\cal {K}$ is the Kernel of $E^{(1)}$ and $\cal {M}$ is its complement, i.e.,
\br
 {\cal K} = \{x \in \hat \lie , \lb x, E^{(1)}\rb =0 \},
\er
such that
\br
\lb  {\cal K} , {\cal K} \rb \subset  {\cal K}, \qquad
\lb  {\cal K} , {\cal M} \rb \subset  {\cal M}, \qquad
\lb  {\cal M} , {\cal M} \rb \subset  {\cal K}.
\nonu
\er
The equations of motion are  determined by solving the zero curvature equation 
\br
\lb \pa_x + E^{(1)} + A_0, \pa_{t_N} + D^{(N)} + D^{(N-1)} + 
\cdots +D^{(0)} \rb = 0, \label{5}
\er
The  solution of eq. ( \ref{5}) may be  systematically constructed  by considering $D^{(j)} \in \lie_j$ and $A_0 \in  {\cal {M}}_0 , \;\;  {\cal {M}}_0\in \lie_0$  
and  can be decomposed according to the graded structure as 
\br
\lb E, D^{(N)}\rb &=0& \label{6.1} \\
\lb E, D^{(N-1)}\rb + \lb A_0, D^{(N)}\rb +\pa_x D^{(N)} &=0& \nonu \\
\vdots &=& \vdots \nonu \\
\lb A_0, D^{(0)}\rb + \pa_x D^{(0)} - \pa_{t_{N}}A_0 &=&0. 
\label{6}
\er
The unknown $D^{(j)}$'s can be solved starting from the highest to the lowest grade projections as functionals of $A_0 $ and its $x-$ derivatives.  
Notice that, in particular   the highest grade equation, namely $\lb E, D^{(N)}\rb =0$  implies $D^{(N)} \in {\cal K}$.  If we  consider the  fields of 
the theory to parametrize $A_0 \in {{\cal M}}_0$, 
the  equations of motion are obtained form the zero grade component  (\ref{6}). 

We shall now work  with an explicit example of the {\it  mkdV hierarchy}  based upon the $\hat \lie = \hat {sl}(2)$ affine algebra, 
\br
\lb h^{(m)}, E_{\pm \a}^{(n)} \rb = \pm 2 E_{\pm \a}^{(n)}, \qquad \lb E_{ \a}^{(m)},E_{- \a}^{(n)}\rb = h^{(m+n)}
\label{1}
\er 
 The grading operator is $Q =2 \l {{{d}\o {d \l}}} + {1\o 2} h$ and decomposes the affine algebra $\hat {\lie} $ into even and odd graded subspaces 
\br
\lie_{2m} &=& \{ h^{(m)} = \l^m h\}, \nonu \\
\lie_{2m+1} &=& \{\l^m\(E_{\a} + \l E_{-\a}\), \; \l^m\(E_{\a} - \l E_{-\a}\)\}
\label{2} 
\er
for $m=0, \pm 1, \pm 2, \cdots$ and $[\lie_a, \lie_b] \subset \lie_{a+b}$.
The integrable hierarchy is then specified by a choice of a semi-simple  element $E = E^{(1)}$, where
\br
E^{(2n+1)} = \l^n \(E_{\a} + \l E_{-\a}\) 
\label{3}
 \er
 and $A_0 = v(x,t_n)h^{(0)}$.  
 The Kernel of $E^{(1)} $ is therefore given by 
\br
 {\cal K} = {\cal K}_{2n+1} =  \{ \l^n\(E_{\a} + \l E_{-\a}\)\} \label{3.5} 
 \er
 and  has grade $2n+1$.  It thus follows from (\ref{6.1}) that the highest grade  component  of $D^{(N)}$ has grade $N=2n+1$.
 The component  within the ${\cal M}$ of the 
zero grade projection of (\ref{6}) leads to the  evolution equations according to time $t=t_{2n+1}$.  Notice that $D^{(0)} $ lies  within the Cartan subalgebra and hence $[A_0, D^{(0)} ] =0$.  The equations of motion  are then  simplified to $\pa_{t_{2n+1}} A_0 = \pa_x D^{(0)}$,
Examples are, 
\br
n=1 \qquad 4\pa_{t_3} v&=& \pa_x \(\pa_x^2v -2 v^3 \)\qquad  mKdV  \nonu \\
\label{t3} \\
\nonu \\
n=2 \qquad 16\pa_{t_5}v  &=& \pa_x \(   \pa_x^4v -10v^2(\pa_x^2 v) -10v(\pa_x v)^2+6v^5 \), \nonu \\
\label{t5} \\
\nonu \\
n=3 \qquad 64\pa_{t_7}v &=& \pa_x \(  \partial_{x}^6v-70 (\partial_{x}v)^{2}(\partial_{x}^2v)-42v(\partial_{x}^2v)^{2}-56 v(\partial_{x}v)(\partial_{x}^3v)\)\nonu \\  
&-&\pa_x \(14 v^{2}\pa_x^4v -140 v^{3}(\partial_{x}v)^{2}-70 v^{4}(\partial_{x}^2v)+20v^{7}  \)\nonu \\ 
\label{t7} \\
  \cdots etc \nonu 
\er

For the negative mKdV sub-hierarchy let us propose the following form for the zero curvature representation 
\br
\lb \pa_x + E^{(1)} + A_0, \pa_{t_{-N}} + D^{(-N)} + D^{(-N+1)} 
+\cdots +D^{(-1)} \rb = 0.
\label{7}
\er
Differently  from the positive hierarchy case, the lowest grade  projection now yields,
\br
\pa_x D^{(-N)} + \lb A_0, D^{(-N)}\rb = 0,
\nonu
\er
 a nonlocal equation for $D^{(-N)}$.   Having solved for $D^{(-N)}$,  the second lowest projection of grade $-N+1$, leads to
 \br
\pa_x D^{(-N+1)} + \lb A_0, D^{(-N+1)} \rb + \lb E^{(1)},D^{(-N)} \rb = 0
\nonu
\er
which determines $D^{(-N+1)}$.  The proccess follows recursively until we reach the zero grade projection 
\br
\pa_{t_{-N}} A_0  - \lb  E^{(1)},D^{(-1)}\rb  = 0
\label{8}
\er
which yields the evolution equation for field $A_0$ according to time $t=t_{-N}$
 Notice that in this case there is no condition upon $N$.
 
 The simplest example is to take $N=1$ when the zero curvature decomposes into 
 \br
\pa_x D^{(-1)} + \lb A_0, D^{(-1)}\rb & = & 0, \nonu \\
\pa_{t_{-1}} A_0  - \lb  E^{(1)},D^{(-1)}\rb  & = & 0.
\label{9}
\er
In order to solve the first equation, we define the zero grade group element $B = \exp \( \lie_0 \) $ and  define
 \br
D^{(-1)} = B E^{(-1)} B^{-1}, \qquad A_0 = -\pa_x B B^{-1}, 
\label{10}
\er
Under such parametrization  the second eqn. (\ref{10}) becomes the well known (relativistic ) Leznov-Saveliev equation,
\br
\pa_{t_{-1}} \(\pa_x B B^{-1}\) + \lb  E^{(1)}, B E^{(-1)}B^{-1} \rb 
\label{11}
\er
which for $\hat{sl}(2)$ with principal gradation 
$Q= 2\l {{d}\o {d\l}} + {1\o 2} h$, yields the sinh-Gordon equation
\br
\pa_{t_{-1}} \pa_x \phi  = e^{2\phi} - e^{-2\phi}, \qquad  B= e^{-\phi h}.
\label{sg}
\er
where $t_{-1} = z, x = \bar z,  A_0 = v h \equiv  \pa_x \phi h$. 

 For higher values of $N=2,3,4,\cdots$ we found
 \br
 \pa_{t_{-2}}\pa_x \phi &=& 4e^{-2\phi} d^{-1} e^{2\phi} + 4e^{2\phi} d^{-1} e^{-2\phi} 
 \label{tm2}
 \er
 \br
 \pa_{t_{-3}}\pa_x \phi &=& 4e^{-2\phi} d^{-1}\( e^{2\phi}d^{-1} (\sinh 2\phi)\) +4 e^{2\phi} d^{-1}\( e^{-2\phi}d^{-1} (\sinh 2\phi)\)
 \label{tm3}
 \er
 \br
 \pa_{t_{-4}}\pa_x \phi &=& 4 e^{-2\phi} d^{-1} \( e^{2\phi}d^{-1} ( e^{-2\phi} d^{-1} e^{2\phi} + e^{2\phi} d^{-1} e^{-2\phi} ) \) \nonu \\
 &+&4 e^{2\phi} d^{-1} \(e^{-2\phi}d^{-1}(e^{-2\phi} d^{-1} e^{2\phi} + e^{2\phi} d^{-1} e^{-2\phi})\)
 \label{tm4}
 \er
 \br
 \pa_{{t_{-5}}}\pa_x \phi &=& 8 e^{-2\phi} d^{-1} \(e^{2\phi} d^{-1}\( e^{-2\phi} d^{-1}\( e^{2\phi}d^{-1} (\sinh 2\phi)\) + e^{2\phi} d^{-1}\( e^{-2\phi}d^{-1} (\sinh 2\phi)\)\)\) \nonu \\
 &+& 8 e^{2\phi} d^{-1} \(e^{-2\phi} d^{-1}\(e^{-2\phi} d^{-1}\( e^{2\phi}d^{-1} (\sinh 2\phi)\) + e^{2\phi} d^{-1}\( e^{-2\phi}d^{-1} (\sinh 2\phi)\)\)\), \nonu 
 \\
 \label{tm5}
 \er
 where $d^{-1}f = \int^x f(y)dy$.  For $N=2$  eqn. (\ref{tm2}) was  derived in \cite{qiao} using recurssion operators.
 
 We now consider  soliton solutions  for the entire hierarchy.  The general algebraic structure  of the zero curvature representation  yields a general method for constructing  soliton solutions based on the fact that $v=0$ ( and/or $\phi=0$) is the vacuum solution for all positive and negative odd sub-hierarchies, i.e., eqns. (\ref{t3})-(\ref{t7}) and (\ref{sg}), (\ref{tm3}), (\ref{tm5}), $\cdots $, etc. \footnote{For negative even sub-hierarchy, the vacuun solution is obtained for $v=v_0 \neq 0$ and the dressing  method works equally well generating  general formulae for multisoliton solutions  but with a deformation  parameter $v_0$, see for instance \cite{nos}}
 The idea of the dressing method is to map the {\it trivial vacuum }   into an nontrivial   configuration by gauge transformation, i.e,
 \br
 \pa_x + {A_0} & = & ({\Theta_{\pm }})^{-1} \(\pa_x + E^{(1)} + A_0^{vac}\) \Theta_{\pm },  \nonu \\
 \nonu \\
\pa_{t_k} + D^{(k)} + \cdots + D^{(0)} & = & ({\Theta_{\pm }})^{-1} \( \pa_{t_k} + D^{k)}_{ vac}+ \cdots + D^{(0)}_{vac} \) \Theta_{\pm } \nonu  
\er
where $\Theta_{\pm}$ are group elements of the form
\br
 \nonu \\
\Theta_{-}^{-1}=e^{p(-1)}e^{p(-2)} \ldots\: , \qquad 
\Theta_{+}^{-1}=e^{q(0)}e^{q(1)}e^{q(2)} \ldots , \nonu \\
\nonu 
\er
 $p^{(-i)}$ and $q^{(i)}$ are linear combinations of grade 
$(-i)$ and $(i)$ generators respectively.  

It thus follows that  one and two soliton solutions  for the mKdV hierarchy with $A_0^{vac} = 0$ can be written as
\br
\phi_{1-sol} &=& ln \({{1-a_1\rho_1} \over {1+a_1\rho_1}}\) \nonu \\
\phi_{2-sol} &=& ln \({{1-a_1\rho_1 -a_2\rho_2+ a_1a_2 a_{12}\rho_1\rho_2} \over {1+a_1\rho_1+a_2\rho_2+ a_1a_2a_{12}\rho_1\rho_2}}\) \label{sol} \\
\vdots &=&\vdots \nonu
\er
$a_{12} =\({{ \k_1 - \k_2}\over {\k_1 + \k_2   }}\)^2  $
and 
\br
\rho_i\(\k_i\)=\exp 2 (k_i x + t_a (k_i)^{a}).  \label{rho} 
\er
These are solutions for all  equations  of the positive and negative odd hierarchies, i.e., eqns. (\ref{t3})-(\ref{t7}) and (\ref{sg}), (\ref{tm3}), (\ref{tm5}), $\cdots $, 
for $a= 3,5,7, \cdots$ and $a=-1, -3, -5, \cdots $ respectively.  

It is clear that $ \phi_{vac} = \phi_0 = 0$ do not satisfy eqns. (\ref{tm2}) or (\ref{tm4}). It follows that the   negative even cases, (\ref{tm2}), (\ref{tm4}), etc,  do not admit  zero vacuum solution, i.e.  $ v_{vac}=  0.$ is not  a solution.    The soliton solutions (\ref{sol})  have to be modified accordingly to $ v_{vac} = v_0 \neq 0$, (see \cite{nos}) and these cases shall be discussed elsewhere.

\section{Backlund Transformation}

In this section we start by noticing  that the zero curvature representation (\ref{5}) or (\ref{7}) of the form 
\br \lb \pa_x + A_x, \pa_t + A_t \rb = 0 
\label{zcc}
\er
are invariant under gauge transformations of the type
\br
A_{\mu} (\phi, \pa_x \phi, \cdots ) \rightarrow \tilde {A}_{\mu}= U^{-1} A_{\mu} U + U^{-1} \pa_{\mu}U
\label{gt}
\er
If we now choose $U(\phi_1, \phi_2)$ such that  it maps one field configuration $\phi_1$ into another field configuration $\phi_2$ preserving the equations of motion (i.e., zero curvature (\ref{zcc})) see \cite{thiago},
\br
U A_{\mu}(\phi_1)  = A_{\mu}(\phi_2) U + \pa_{\mu} U
\label{back}
\er
If we now take $A_{\mu} = A_x = E^{(1)} + A_0$ which is {\it common to all members} of the hierarchy,
we find that 
\br
U=\left[ \begin{array}{cc}  1 & -{{\b}\over {2\l}}e^{-(\phi_1+\phi_2)} \\ -{{\b}\over {2}}e^{(\phi_1+\phi_2)} & 1 \end{array} \right]
\label{gt}
\er
satisfies (\ref{back}) provided
\br
\pa_x \( \phi_1 - \phi_2\) = -\b sinh \( \phi_1 + \phi_2 \).
\label{xbl-sg}
\er 
For the sinh-Gordon  model, the equations of motion (\ref{sg}) are satisfied if we further introduce the time component of the Backlund transformation,
\br
\pa_{t_{-1}} \( \phi_1 + \phi_2\) = {{4}\over {\b}} sinh \( \phi_2 - \phi_1 \).
\label{tbl-sg}
\er 
where $\phi_a, a=1,2$ satisfy the sinh-Gordon eqn, $ \pa_{t_1} \pa_x \phi_a = {{1}\over {2}} sinh \phi_a$. 
The gauge transformation (\ref{gt})  leads to the Backlund transformation for the negative odd sub-hierarchy.  Consider first the $t=t_{-3}$ evolution equation (\ref{tm3}) where
\br A_{t_{-3}} = D^{(-3)} + D^{(-2)} + D^{(-1)}
\label{atm3}
\er
where
\br
D^{(-3)}(\phi) &=& -a\( e^{-2\phi}E_{\a}^{(-1)} + e^{2\phi} E_{-\a}^{(-1)}\), \qquad \qquad
D^{(-2)}(\phi) = 2a I(\phi) h^{(-1)} \nonu \\
D^{(-1)}(\phi) &=& -4a\( e^{-2\phi} \int^{x} e^{2\phi} I(\phi) E_{\a}^{(-1)} +e^{2\phi} \int^{x} e^{-2\phi} I(\phi) E_{-\a}^{(0)}\) \label{dtm3}
\er
where $I(\phi_i) = \int^x \sinh (2\phi_i), \;\; i=1,2$.  Inserting  the gauge transformation $U$ (\ref{gt}) into (\ref{back}) for $A_{\mu} = A_{t_{-3}}$  given in (\ref{atm3}) and (\ref{dtm3}) we find the Backlund transformation
\br
\pa_{t_{-3}}(\phi_1+\phi_2) &=& {{8}\over {\b}}e^{\phi_1-\phi_2}\int^x e^{2\phi_2}I(\phi_2)  - {{8}\over {\b}}e^{-\phi_1+\phi_2}\int^x e^{2\phi_1}I(\phi_1), \nonu \\
\label{btm3}
\er
together with the subsidiary conditions
\br
I(\phi_2) - I(\phi_1) &=& \b e^{\phi_1-\phi_2} \int^x \(e^{-2\phi_1} I(\phi_1)+ e^{2\phi_2}I(\phi_2)\)\nonu \\
I(\phi_2) + I(\phi_1) &=& {{2}\over {\b}}\sinh (\phi_2 -\phi_1)
\label{bl3}
\er

The very same argument follows for $t=t_{-5}$ where  $A_{t_{-5}} = D^{(-5)} + D^{(-4)}+D^{(-3)} + D^{(-2)} + D^{(-1)}$. Here (\ref{gt}) into (\ref{back}) yields,
\br 
\pa_{t_{-5}}(\phi_1+\phi_2) &=& {{16}\over {\b}} e^{\phi_1-\phi_2} \int^x e^{2\phi_2} W(\phi_2) - {{16}\over {\b}} e^{-\phi_1+\phi_2} \int^x e^{-2\phi_2} W(\phi_2)
\er
where $W(\phi_i) = \int^x \(e^{-2\phi_i} \int^{y}e^{2\phi_i}I(\phi_i)\)dy + \int^x \(e^{2\phi_i} \int^{y}e^{-2\phi_i}I(\phi_i) \)dy$
together with  the subsidiary conditions
\br
W(\phi_2) - W(\phi_1) &=& \b e^{\phi_1-\phi_2} \(\int^x e^{-2\phi_1} W(\phi_1)  + e^{2\phi_2}W(\phi_2)\)\nonu \\
I(\phi_2) - I(\phi_1) &=& \b e^{\phi_1-\phi_2} \(\int^x e^{-2\phi_1} I(\phi_1)  + e^{2\phi_2}I(\phi_2)\)\nonu \\
I(\phi_2) + I(\phi_1) &=& {{2}\over {\b}} \sinh (\phi_2-\phi_1).
\label{bl5}
\er
Other subsidiary relations  are obtained from (\ref{bl3}) and (\ref{bl5}) by  replacing $\phi_i \rightarrow -\phi_i$.

The key observation that  allows us to extend such Backlund transformation to other positive higher grade members of the hierarchy (\ref{t3} - \ref{t7}), etc is to notice that  the zero grade component of  equation (\ref{6}) is trivially solved by parametrizing $A_0 =  -\pa_x B B^{-1}$ and $D^{(0)} = d_0h^{(0)} = -\pa_{t_{2n+1}} B B^{-1}$.
On the other hand by solving grade by grade the zero curvature eqn. (\ref{6.1})-(\ref{6})  we find  explicit expressions  (\ref{b3})-(\ref{b7}) for $D^{(0)}$. We define  then the multi-time evolution  for the field $v(x,t_N) = \pa_x \phi(x, t_N)$ to be
\br
n=1  \qquad   \pa_{t_3}\phi(x, t) & \equiv &  d_0  ={{1}\o {4}} \pa_x^3 \phi - {{1}\over {2}} ( \pa_x \phi )^3  
\label{c3a} \\
n=2  \qquad \pa_{t_5}\phi(x, t) & \equiv &  d_0  ={{1\o {16}}}\pa_x^5 \phi  -{{5}\o {8}} (\pa_x \phi)^2 \pa_x^3\phi -{{5}\o {8}}
\pa_x \phi (\pa_x^2\phi)^2 +{{3}\o {8}} (\pa_x \phi)^3,  
\label{c5} \\
n=3 \qquad \pa_{t_7}\phi(x, t) & \equiv &  d_0 = \frac{1}{64} \partial_{x}^7\phi-\frac{35}{32}(\partial_{x}^2\phi)^{2}(\partial_{x}^3\phi)-\frac{21}{32}(\pa_x\phi)(\partial_{x}^3\phi)^{2}-\frac{7}{8}(\pa_x\phi)(\partial_{x}^2\phi)(\partial_{x}^4\phi)\nonu \\  &-&\frac{7}{32}(\pa_x \phi)^{2}(\partial_{x}^5\phi)+\frac{35}{16}(\pa_x \phi)^{3}(\partial_{x}^2\phi)^{2}+\frac{35}{32}(\pa_x \phi)^{4}(\partial_{x}^3\phi)-\frac{5}{16}\pa_x^{8} \phi \\
\label{c7} 
  \cdots etc \nonu 
\er

It  follows that the Backlund transformation  for the time component  $t_3$ may be derived from  the above eqn. (39) by  considering 
\br
4\pa_{t_3}(\phi_1 - \phi_2)= \pa_x^3 \phi_1 -\pa_x^3 \phi_2  -2  (\pa_x \phi_1 )^3 +2 (\pa_x \phi_2)^3 , \nonu 
\er
 Eliminating  $\pa_x \phi_2$ from $x-$  component of Backlund transf. for Sinh-Gordon, i.e., $\pa_x (\phi_1 - \phi_2) = -\b sinh(\phi_1 +\phi_2)$,
 we find
\br
4(\partial_{t_3}\phi_{2}-\partial_{t_3}\phi_{1})&= & \b(\partial^2_{x}\phi_{1}+\partial^2_{x}\phi_{2})\cosh(\phi_{1}+\phi_{2})
-\frac{\b}{2}(\partial_{x}\phi_{1}+\partial_{x}\phi_{2})^{2}\sinh(\phi_{1}+\phi_{2}) \nonumber \\ &-&\frac{\b^{3}}{2}\sinh^{3}(\phi_{1}+\phi_{2}).
\label{b3}
\er
which is in agreement with \cite{rogers}.

Analogously the same follows from (\ref{c5}) and (\ref{c7})  for $t_5$ and $t_7$ respectively, yielding
\br
16\partial_{t_5}(\phi_{2}-\phi_{1})&= & 2\b \partial^4_{x}\phi_{1}\cosh(\phi_{1}+\phi_{2})
-4\b \partial_{x}\phi_{1}(\partial^3_{x}\phi_{1})\sinh(\phi_{1}+  \phi_{2})\nonu \\
&-&12\b (\partial_{x}\phi_{1})^{2}\partial^2_{x}\phi_{1}\cosh(\phi_{1}+\phi_{2}) 
+2\b (\partial^2_{x}\phi_{1})^{2}\sinh(\phi_{1}+\phi_{2})\nonu \\
&+& 6\b (\partial_{x}\phi_{1})^{4}\sinh(\phi_{1}+\phi_{2})
- 4\b^{2}(\partial_{x}\phi_{1})^{3}+2\b^{2}\partial^3_{x}\phi_{1}\nonu \\
&-&   2\b^{3}(\partial_{x}\phi_{1})^{2}\sinh(\phi_{1}+\phi_{2})
+2\b^{3}\partial^2_{x}\phi_{1}\cosh(\phi_{1}+\phi_{2}) \nonumber  \\
&+&2 \b^4\pa_x \phi_1 + \b^{5}\sinh(\phi_{1}+\phi_{2})\label{b5}
\er
and 
 
\br
64(\partial_{t_7}\phi_{2}-\partial_{t_7}\phi_{1})&=&-20\b(\partial_{x}\phi_{1})^{6}\sinh(\phi_{12})
+20\b \pa_x(\phi_1)^2(\pa_x^2\phi_1)^2 \sinh(\phi_{12}) \nonu \\
&-&20 \b (\pa_x^2\phi_1)^3\cosh(\phi_{12}) 
+ 40 \b (\pa_x\phi_1)^3 \pa_x^3\phi_1 \sinh(\phi_{12}) \nonu \\
&-&80\b \pa_x\phi_1 \pa_x^2\phi_1 \pa_x^3\phi_1 \cosh(\phi_{12}) 
-2 \b (\pa_x^3\phi_1)^2\sinh(\phi_{12}) \nonu \\
&-&20\b (\pa_x\phi_1)^2 \pa_x^4\phi_1 \cosh(\phi_{12}) 
+4 \b \pa_x^2 \phi_1 \pa_x^4 \phi_1 \sinh(\phi_{12}) \nonu \\
&+&60 \b (\pa_x \phi_1)^4 \pa_x^2 \phi_1 \cosh(\phi_{12}) 
-4 \b \pa_x\phi_1 \pa_x^5 \phi_1 \sinh(\phi_{12}) \nonu \\
&+& 2 \b \pa_x^6\phi_1 \cosh(\phi_{12}) 
+12 \b^2 (\pa_x\phi_1)^5 -20 \b^2 \pa_x \phi_1(\pa_x^2\phi_1)^2  \nonu \\
&-&20\b^2 (\pa_x\phi_1)^2 \pa_x^3 \phi_1 +2\b^2 \pa^5_x \phi_1 
-12\b^3(\pa_x\phi_1)^4\sinh(\phi_{12}) \nonu \\
&-&12 \b^3 (\pa_x\phi_1)^2 \pa_x^2 \phi_1 \cosh(\phi_{12}) 
+2\b^3(\pa_x^2\phi_1)^2\sinh(\phi_{12}) \nonu \\
&-& 4 \b^3 \pa_x\phi_1 \pa^3_x \phi_1\sinh(\phi_{12}) 
+2\b^3 \pa_x^4 \phi_1 \cosh(\phi_{12}) \nonu \\
&-&4 \b^4 (\pa_x \phi_1)^3+2 \b^4 \pa^3_x \phi_1 
-2 \b^5(\pa_x \phi_1)^2\sinh(\phi_{12}) \nonu \\
&+&2 \b^5 \pa_x^2 \phi_1 \cosh(\phi_{12})
+2 \b^6 \pa_x \phi_1 + \b^7 \sinh(\phi_{12}) \label{b7}
\er
where $\phi_{12} \equiv \phi_1 + \phi_2$.


\section{Examples }
In this section we shall consider few solutions for the Backlund  solutions for (\ref{t3})-(\ref{t7}) and (\ref{sg}), (\ref{tm3}) and (\ref{tm5}).
\subsection{ Vacuum - 1-soliton}

Let 
\br \phi_1 = \phi_{vac}= 0, \qquad \phi_2 = \phi_{1-sol}=\ln \({{1+R\rho}\over {1-R\rho}}\), \qquad \rho = e^{2kx+2k^Nt_N}
\label{vac-sol}
\er
and $R$ is a constant.
It becomes clear that the Backlund equations (\ref{b3})-(\ref{b7}) 
are satisfied by (\ref{vac-sol}) for $\b = 2k$.
 
\subsection{ 1-soliton - 1-soliton}
\br 
\phi_1 = \phi_{1-sol}(k_1) = \ln \({{1+R_1\rho_1}\over {1-R_1\rho_1}}\), \qquad \phi_2 = \phi_{1-sol}(k_2) = \ln \({{1+R_2\rho_2}\over {1-R_2\rho_2}}\), \qquad \rho_i = e^{2k_ix+2k_i^Nt_N}
\label{sol-sol}
\er
and $R_i= R_i(k_i)$.  Backlund solutions given in (\ref{sol-sol}) satisfy  the Backlund equations (\ref{b3})-(\ref{b7})  for 
\br
k_1 = k_2 = k, \qquad R_2 = \({{2k + \b}\over{2k-\b}}\)R_1.
\er
respectively for $N=3,5$ and $7$.  It was verified that the same also satisfy eqns. (\ref{sg}), (\ref{tm3}) and (\ref{tm5}) with $N=-1, -3$ and $-5$ respectively.
\subsection{ 1-soliton - 2-soliton}
\br 
\phi_1 = \phi_{1-sol} = \ln \({{1+\rho_1}\over {1-\rho_1}}\), \qquad \phi_2 = \phi_{2-sol} = \ln \({{1+\d (\rho_1-\rho_2)-\rho_1\rho_2}\over {1-\d (\rho_1-\rho_2)-\rho_1 \rho_2}}\), 
\label{sol-2sol}
\er
where $\d = {{k_1+k_2}\over {k_1-k_2}}$.

We have verified that the Backlund  equations for the higher members of the hierarchy, namely,(\ref{b3})-(\ref{b7}) are satisfied by (\ref{sol-2sol}) for $N=3,5$ and $7$.  Also, the permutability theorem, which establishes the equality  of the 2-soliton solution obtained from  vacuum to 1-soliton solution for $\b = 2k_1$ and subsequently this 1-soliton to 2-soliton solutions and $\b=2k_2$.  On the other hand the very same 2-soliton solution is obtained from   the vacuum to 1-soliton solution for $\b = 2k_2$ and subsequently this 1-soliton to 2-soliton solutions and $\b=2k_1$.  
For technical reasons we  were unable to verify this case for  the negative odd equations (\ref{sg}), (\ref{tm3}) and (\ref{tm5}).

\begin{figure}[h]
\centering
\includegraphics[width=10cm]{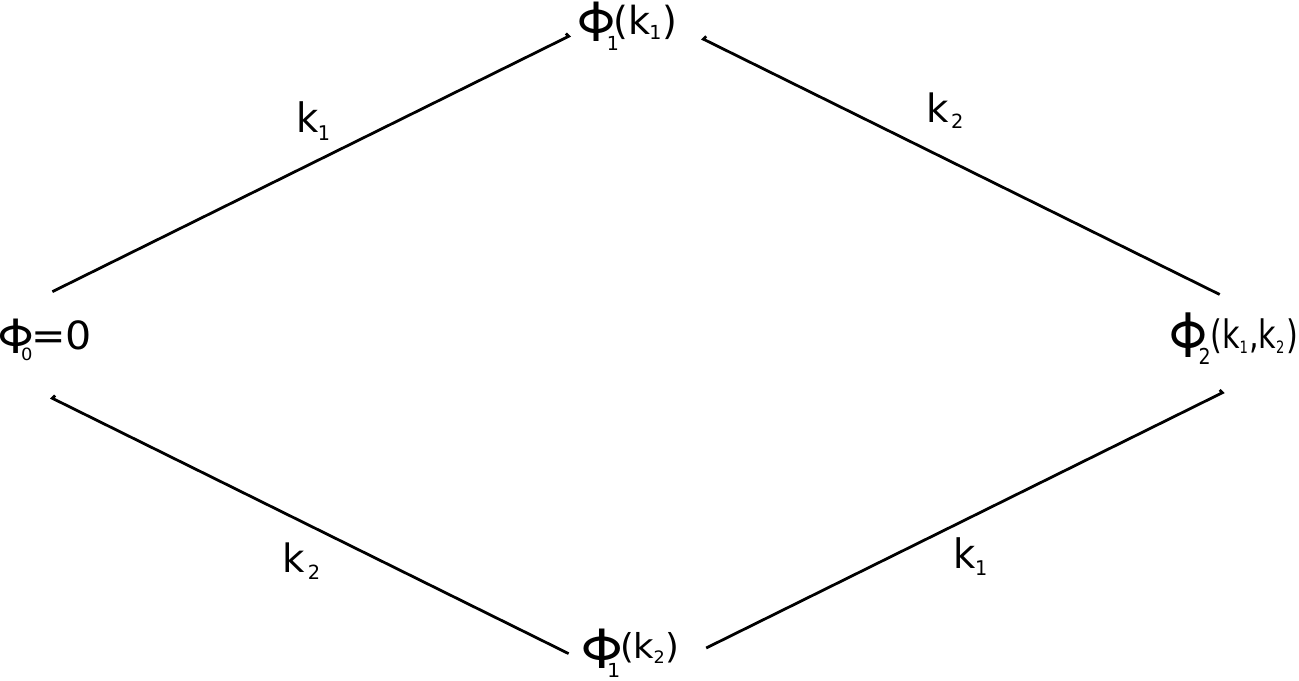}
\caption{ Permutability theorem for 2-solitons  solution }
\label{1}
\end{figure}


\section{Conclusions and Further Remarks}

The construction of a set of non-linear integrable equations of motion were defined in terms of a zero curvature representation and an  affine Lie algebra.  Explicitly, we have considered the mKdV hierarchy  for positive and negative odd graded time evolutions.
A class of soliton solutions, labeled  according  to different graded time evolutions, were  constructed in a universal manner to all  members of the hierarchy.    
These were derived, by the dressing method from the trivial vacuum solution , i.e. $v_{vac}=0$.

We have extended the Backlund transformation  of the sinh-Gordon model to other  higher and lower graded members of the mKdV hierarchy.  We have  shown that the spatial component of the Backlund transformation  is common to all members of the hierarchy  and is  a direct consequence of the  common Lax operator.
These equations are verified to be satisfied  by the same
 pair of solutions   for the first few members of the hierarchy.
 
 For the negative  even graded equations the trivial vacuum is observed {\it not} to be solution of the equations of motion and a deformation  of the dressing method need to be employed (see \cite{nos}).
 
We are extending  the same arguments from the $N=1$ super sinh-Gordon model to the  supersymmetric   mKdV hierarchy \cite{ymai}.

So far,  our arguments were based upon known  spatial component of the Backlund transformation of the Sinh-Gordon model.  Such relation defines the so called type I Backlund transformation.   The extension  to type II Backlund transformation \cite{corr} , \cite{thiago} is a subject  for future investigation.

\vskip 10pt \noindent
{\bf Acknowledgements} \\
We  would like to thank  Capes, CNPq and Fapesp for support.


\section*{References}
\begin{thebibliography}{9}
\bibitem{nissimov} Aratyn, H,  Gomes, JF,   Nissimov, E,  Pacheva, S and  Zimerman, AH (2000), " Symmetry flows, conservation laws
and dressing approach to the integrable models.", NATO Advanced Research Workshop on Integrable
Hierarchies and Modern Physical Theories, Chicago, Illinois, p., 243-275, nlin/0012042
\bibitem{1} Chodos, A (1980)
  Phys. Rev.  D { 21}, 2818. 
  \bibitem{2} Tracy, CA and  Widom, H (1976) Comumm. Math. Phys. { 179},1
    \bibitem{jpa} Aratyn, H,   Ferreira, LA,  Gomes, JF, and  Zimerman, AH (2000) J. of Physics { A33},L331, 
 nlin/0007002

\bibitem{sotkov} 	 Gomes, JF,  Gueuvoghlanian, EP,  Sotkov, GM, and   Zimerman, AH,(2001) 
 Nucl.Phys. B606  441, hep-th/0007169 

\bibitem{iraida}  Cabrera-Carnero, I,   Gomes, JF,  Sotkov, GM, and  Zimerman, AH, (2002) 
 Nucl.Phys. B634  433, hep-th/0201047 

\bibitem{2bh} 	 Gomes, JF,  Sotkov,GM, and   Zimerman, AH (2005) 
 Nucl.Phys. B714 179,  hep-th/0405182 
 
 \bibitem{nos}  Gomes, JF, Starvaggi Fran\c ca, G,   de Melo, GR,  and Zimerman, AH, (2009) J. of Physics { A42},445204, arXiv:0906.5579 

 
 \bibitem{corrigan1}  	
 Bowcock, P,   Corrigan, E and   Zambon, C (2004) Int.J.Mod.Phys. A19, S2  82, hep-th/0305022 

\bibitem{corrigan-toda} 	
 Bowcock, P,  Corrigan, E and   Zambon, C (2004) JHEP 0401  056, 
 hep-th/0401020 
 
 \bibitem{corrigan-nr}  Corrigan, E and   Zambon, C (2006)
 Nonlinearity 19  1447-1469, nlin/0512038
 


\bibitem{wadati1}  Kodama, Y and  Wadati, M (1976) Progr. of Theor. Phys. { 56}, 1740
  \bibitem{villani}   Villani, A and  Zimerman, AH (1977) Rev. Bras. de Fisica, { 7}, 649
  
 \bibitem{rogers}  Rogers, C, and  Shadwick, WF, (1982)  Backlund Transformation and their Application, Academic Press, New York 
 
 
  \bibitem{ymai} Gomes, JF,   Ymai, LH and   Zimerman, AH (2006),  Phys. Lett. A359:630,  arXiv:hep-th/0607107 
  \bibitem{corr} Corrigan, E,  and  Zambon, C, (2009), J. of Phys. A42,  475203,  arXiv:0908.3126
  
  
  
  
 \bibitem{qiao}  Qiao, Z and  Strampp, W, (2002), Physica A313,365 
  
  \bibitem{thiago}  Aguirre, AR,   Araujo, TR,  Gomes, JF and  Zimerman, AH, (2011),  
   Journal of High Energy Physics,  12, 56,  arXiv:1110.1589
    
\end{thebibliography}
\end{document}